\documentclass[runningheads]{svmult}

\usepackage{makeidx}   
\usepackage{graphicx}  
\usepackage{subeqnar}  
\usepackage{multicol}  
\usepackage{physprbb}  
\makeindex             

\begin{document}
\title*{\bf A Gamma-Ray Bursts' Fluence-Duration Correlation}

\toctitle{A Gamma-Ray Bursts' Fluence-Duration 
\protect\newline Correlation}
%

\titlerunning{A Gamma-Ray Bursts' Fluence-Duration Correlation}

\author{Istv\'an Horv\'ath\inst{1}
\and Lajos G. Bal\'azs\inst{2}
\and Peter M\'esz\'aros\inst{3}
\and Zsolt Bagoly\inst{4}
\and Attila M\'esz\'aros\inst{5}}
\authorrunning{Istv\'an Horv\'ath et al.}
%

\institute{Dept. of Physics, Bolyai Military University, Budapest, 
Box 12, H-1456, Hungary
\and Konkoly Observatory, Budapest, Box 67, H-1525, Hungary
\and Department of Astronomy and Astrophysics, Pennsylvania State 
University, 525 Davey Lab. University Park, PA 16802, USA
\and Lab. for Information Technology, E\"{o}tv\"{o}s Univ., 
  P\'azm\'any s. 1/A, H-1518, Hungary
\and Astron. Inst. Charles Univ., 180 00 Prague 8,
 V Hole\v sovi\v ck\'ach 2, Czech Republic }

\maketitle              

\begin{abstract}

We present an analysis indicating that there is a correlation between
the fluences and the durations of gamma-ray bursts, and provide arguments
that this reflects a correlation between the total emitted energies and
the intrinsic durations.  For the short (long) bursts the total emitted
energies are roughly proportional to the first (second) power of the
intrinsic duration. This difference in the energy-duration relationship
is statistically significant, and may provide an interesting constraint
on models aiming to explain the short and long  gamma-ray bursts.

\end{abstract}

\section{Introduction}

The gamma-ray bursts (GRBs) measured with the BATSE instrument on the 
Compton Gamma-Ray Observatory are usually characterized by 
9 observational quantities (2 durations, 4 fluences, 3 peak fluxes) 
\cite{me96}, \cite{pac}, \cite{me00}. 
In a previous paper \cite{bag98} 
we have shown that these 9 quantities 
can be reduced to only two significant independent 
variables (principal components). Here we present a new statistical
analysis of the correlation between these variables and show that
there is a significant difference between the power law exponents 
of long and short bursts. The details of this analysis will be presented
elsewhere \cite{bal01}.

\section{Distributions of Durations and Total Emitted Energies}
\label{sec:durations}

We consider here those GRBs from the current 
BATSE Gamma-Ray Burst Catalog \cite{me00}
 which have measured $T_{90}$ durations and fluences
 ($ F_1, F_2, F_3, F_4$).
Therefore, we are left with $N=1929$ GRBs, all of which have defined 
$T_{90}$ and $F_{tot}(= F_1 + F_2 + F_3 + F_4$), as well as 
peak fluxes $P_{256}$. 

The distribution of the log$T_{90}$ clearly displays 
two peaks reflecting the existence of two groups of GRBs
\cite{k}. 
This bimodal distribution can be fitted by two log-normal 
distributions \cite{hor}.
The fact that the distribution of $T_{90}$ within a subclass is log-normal
has important consequences. Let us denote the observed duration of a
GRB with $T_{90}$ (which may be subject to cosmological 
time dilatation) and with
$t_{90}$ those measured by a comoving observer (intrinsic duration). 
Then one has $T_{90} = t_{90} f(z)$ 
where $z$ is the redshift, and $f(z)$ measures the time dilatation. 
For the concrete form of $f(z)$ one can take $f(z) = (1+z)^k$, 
where $k=1$ or $k=0.6$, depending on whether energy stretching is included
or not. 

Taking the logarithms of both sides of this equality
one obtains the logarithmic duration as a sum of two independent
stochastic variables. According to a theorem of Cram\'er \cite{cra37}, if 
a variable $\zeta$, which has a Gaussian distribution, is given by a sum of 
two independent variables, i.e. $\zeta = \xi + \eta$, then both 
$\xi$ and $\eta$ have Gaussian distributions. Therefore, from this theorem 
it follows that the Gaussian distributions of $\log T_{90}$, 
confirmed for the two subclasses separately \cite{hor},
implies the same type of distribution for the variables of $\log t_{90}$ and 
of $\log f(z)$. However, unless the space-time geometry has a very 
particular structure, the distribution of $\log f(z)$ cannot be Gaussian. 
This means that the Gaussian nature of the distribution of $\log T_{90}$ 
must be dominated by the distribution of $\log t_{90}$, and the latter 
must then necessarily have a Gaussian distribution. This holds for both 
duration subgroups separately. 
(Note here that several other authors,
e.g. \cite{wp94}, \cite{nor94}, \cite{nor95},
have already suggested, that the distribution of $T_{90}$
reflects predominantly the distribution of $t_{90}$.)

One also has $F_{tot} = (1+z) E_{tot}/(4\pi d_l^2(z)) = c(z) E_{tot}$,
where $d_l$ is the luminosity distance, and $E_{tot}$ is the total
emitted energy. Once there is a log-normal distribution
for $F_{tot}$ (for the two subgroups separately), then the
previous application of Cram\'er theorem is also possible here. 
The existence of this log-normal distribution is not obvious, but may be
shown as follows.

Assume both the short and the long groups have
distributions of the variables $T_{90}$ and $F_{tot}$ which are
log-normal. In this case, it is possible to fit {\it simultaneously} 
the values of $\log F_{tot}$ and $\log T_{90}$ by a single two-dimensional 
("bivariate") normal distribution. This distribution has five parameters 
(two means, two dispersions, and the angle ($\alpha$) between the axis 
$\log T_{90}$ and the semi-major axis of the ``dispersion ellipse"). 
Its standard form can be seen in \cite{tw53} (Chapt. 1.25).
When the $r$-correlation coefficient differs from zero, then 
the semi-major axis of the dispersion ellipse represents a linear relationship 
between $\log T_{90}$ and $\log F_{tot}$ with a slope of $m=\tan \alpha$. This
linear relationship between the logarithmic variables implies a power law
relation of form $F_{tot} = (T_{90})^m$ between the fluence and the 
duration, where $m$ may be different for the two subgroups. Then
a similar relation will exist between
$t_{90}$ and $E_{tot}$.

We obtain the best fit through a maximum likelihood 
estimation (e.g., \cite{KS76}, Vol.2., p.57-58).
From this estimation we obtain
the dependence of the total emitted on the intrinsic duration in form
\begin{equation}
E_{tot} \propto \cases{ (t_{90})^{1}~~;&~~(short bursts); \cr
                        (t_{90})^{2.3}~~;&~~(long bursts). \cr}
\label{eq:ftotpower}
\end{equation}

Several papers discuss the biases in the BATSE values of
$F_{tot}$ and $T_{90}$ (cf. \cite{ep92}, \cite{lamb93}, \cite{lp96}, 
\cite{pl96}, \cite{lp97}, \cite{pac},
\cite{hak00}, \cite{meg00}). 
All types of biases are particularly essential for
faint GRBs. To discuss these effects we provide several different
additional calculations (for more details see \cite{bal01}),
which give the same results.

\section{Conclusion}

The exponent in the power laws differ significantly for the two subclasses
of short ($T_{90} < 2$ s) and long ($T_{90} > 2$ s) bursts.
These new results  may
indicate that two different types of central engines are at work, or
perhaps two different types of progenitor systems are involved. 
While the nature of the progenitors remains so far
indeterminate, our results indicate strongly that the nature of the energy
release process giving rise to the bursts is different between the two
burst classes. 
In the short ones the total energy released is proportional to
the duration, while in the long ones it is proportional roughly to the
square of the duration. This result is completely model-independent, and
provides an interesting constraint on the two types of bursts.

This research was supported in part through
OTKA grants T024027 (L.G.B.), F029461 (I.H.) and T034549,
NASA grant NAG5-2857, Guggenheim Foundation
and Sackler Foundation (P.M.) and
Research Grant J13/98: 113200004 (A.M.).


\begin{thebibliography}{8.}
\addcontentsline{toc}{section}{References}

\bibitem{bag98} Z. Bagoly, A. M\'esz\'aros, I. Horv\'ath,
L.G. Bal\'azs, P. M\'esz\'aros: ApJ \textbf{498}, 342 (1999)
\bibitem{bal01} L.G.
Bal\'azs, P. M\'esz\'aros, Z. Bagoly, I. Horv\'ath, 
A. M\'esz\'aros: astro-ph/0007438
\bibitem{cra37} H. Cram\'er:  \emph{Random variables and
probability distr.} Cambridge Tracts in Maths and
Mathematical Phys. No.36 (Cambridge Univ. Press, Cambridge 1937)
\bibitem{ep92} B. Efron, V. Petrosian:
 ApJ \textbf{339}, 345 (1992)
\bibitem{hak00} J. Hakkila,
C.A. Meegan, G.N. Pendleton, R.S. Mallozzi, D.J. Haglin, R.J. Roiger: 
 In: \emph{Gamma-Ray
Bursts; 5th Huntsville Symp.} ed. by R.M. Kippen, R.S. Mallozzi, G.J.
Fishman (AIP, Melville 2000) pp. 48-52
\bibitem{hor} I. Horv\'ath:  ApJ \textbf{508}, 757 (1998)
\bibitem{KS76} M. Kendall, A. Stuart: \emph{The Advanced Theory
of Statistics} (Griffin, London 1976)
\bibitem{k} C. Kouveliotou et al.:  
ApJ \textbf{413}, L101 (1993)
\bibitem{lamb93} D.Q. Lamb, C. Graziani, I.A. Smith:
 ApJ \textbf{413}, L11 (1993)
\bibitem{lp96} T. Lee, V. Petrosian: ApJ
\textbf{470}, 479 (1996)
\bibitem{lp97} T. Lee, V. Petrosian:  ApJ
\textbf{474}, 37L (1997)
\bibitem{me96} C.A. Meegan et al.: ApJS
\textbf{106}, 65 (3B BATSE Catalog) (1996)
\bibitem{me00}
C.A. Meegan et al.: The BATSE Current Gamma-Ray Burst Catalog (2000)
http://gammaray.msfc.nasa.gov/batse/grb/catalog/current/ 
\bibitem{meg00} C.A.
Meegan, J. Hakkila, A. Johnson, G. Pendleton,
R.S. Mallozzi:  In: \emph{Gamma-Ray
Bursts; 5th Huntsville Symp.} ed. by R.M. Kippen, R.S. Mallozzi, G.J.
Fishman (AIP, Melville 2000) pp. 43--47
\bibitem{nor94} J.P. Norris et al.: ApJ \textbf{424}, 540 (1994)
\bibitem{nor95} J.P. Norris et al.: 
 ApJ \textbf{439}, 542 (1995)
\bibitem{pac} W.S. Paciesas et al.: 
ApJS \textbf{122}, 465 (4B BATSE Catalog) (1999)
\bibitem{pl96} V. Petrosian, T. Lee: ApJ
\textbf{467}, 29L (1996)
\bibitem{tw53} R.J. Trumpler, H.F. Weaver:
 \emph{Statistical Astronomy} 
(University of California Press, Berkeley 1953)
\bibitem{wp94} R.A.M.J. Wijers, B.
Paczy\'nski:  ApJ \textbf{437}, L107 (1994)

\end{thebibliography}
\end{document}